\documentclass{webofc}
\usepackage[varg]{txfonts}   
\usepackage{xcolor}
	\definecolor{darkred}{rgb}{0.5,0,0}
	\definecolor{darkblue}{rgb}{0,0,0.5}
	\definecolor{darkgreen}{rgb}{0.0, 0.5, 0.0}
	 \definecolor{armygreen}{rgb}{0.0, 0.26, 0.15}
	 \definecolor{ams}{rgb}{0.43, 0.5, 0.5}
	 \definecolor{brg}{rgb}{0.0, 0.26, 0.15} 
\usepackage{mdwlist}
\usepackage{tabularx}

\begin{document}
\title{Estimated performance of the TRIUMF ultracold neutron source and electric dipole moment apparatus}
%
%

\author{\firstname{Steve} \lastname{Sidhu}\inst{1,2}\fnsep\thanks{\email{ssidhu@triumf.ca}} \and
        \firstname{Wolfgang} \lastname{Schreyer}\inst{1} \and
        \firstname{Sean} \lastname{Vanbergen}\inst{1,3}\and       
        \firstname{Shinsuke} \lastname{Kawasaki}\inst{4} \and
        \firstname{Ryohei} \lastname{Matsumiya}\inst{1,5} \and
        \firstname{Takahiro} \lastname{Okamura}\inst{1,3}   \and       
        \firstname{Ruediger} \lastname{Picker} \inst{1,2}\fnsep\thanks{\email{rpicker@triumf.ca}} 
        \firstname{} \lastname{for the TUCAN Collaboration}               
}

\institute{TRIUMF, Vancouver, BC 
\and
           Simon Fraser University, Burnaby, BC 
\and
           University of British Columbia, Vancouver, BC 
\and
           KEK, Tsukuba, Ibaraki, Japan
\and
           RCNP, Osaka, Ibaraki, Japan
          }

\abstract{%
  Searches for the permanent electric dipole moment of the neutron (nEDM) provide strong constraints on theories beyond the Standard Model of particle physics.
  The TUCAN collaboration is constructing a source for ultracold neutrons (UCN) and an apparatus to search for the nEDM at TRIUMF, Vancouver, Canada.
  In this work, we estimate that the spallation-driven UCN source based on a superfluid helium converter will provide $(1.38\pm0.02) \times 10^7$ polarized UCN at a density of $217\pm3$~UCN/cm$^3$ to a room-temperature EDM experiment per fill.
  With $(1.43\pm0.02) \times 10^6$ neutrons detected after the Ramsey cycle, the statistical sensitivity for an nEDM search per storage cycle will be $(1.94\pm0.06) \times 10^{-25}\,e$cm (1$\sigma$).
  The goal sensitivity of $10^{-27}\,e$cm (1$\sigma$) can be reached within $281\pm16$ measurement days.
  
}
\maketitle
\section{Introduction}
\label{intro}
The TRIUMF Ultracold Advanced Neutron (TUCAN) Collaboration is building a novel ultracold neutron (UCN) source and an apparatus to search for an electric dipole moment of the neutron (nEDM).
We will outline the design of source and apparatus and estimate their performance. 

The nEDM is a unique observable that signifies time-reversal violation in the neutron.
Assuming the combined charge, parity and time reversal (CPT) operation is symmetric, this also violates CP symmetry.
So far, a non-zero nEDM has not been found.
The current best upper limit is $1.8 \times 10^{-26} \,e$cm (90\% CL)~\cite{Abel:2020gbr}, five orders of magnitude larger than the EDM prediction by the Standard Model of particle physics~\cite{MCKELLAR1987556}.
Beyond-the-Standard-Model (BSM) theories tend to predict additional contributions to the nEDM.
On the one hand, further improving the nEDM upper limit seriously constrains BSM theories.
On the other hand, a discovery above the Standard Model prediction would be an indication of new physics or shed light on the strong CP problem.
The aim of the TUCAN collaboration is to search for an nEDM at a sensitivity at the $10^{-27} e\,$cm level ($1\sigma$).

Similar to other experimental efforts~
\cite{wurm2021panedm, ayres2021design, PhysRevC.97.012501, Tsentalovich:2014mfa}, we will use ultracold neutrons that are characterized by their exceptionally low kinetic energies $E_{\rm kin} < 300 \,$neV. 
The strong interaction of neutrons with materials gives rise to Fermi's pseudopotential $U_{\rm F}$ and can cause reflections under all angles of incidence at a material boundary if $E_{\rm kin} < U_{\rm F}$.
Fermi potentials can be as high as $347\,$neV ($^{58}$Ni).
The interaction of the neutron's magnetic moment $\mu_{\rm n} = -60.3\,$~neV/T with a strong magnetic field $B$ gives rise to similar potentials $U_{\rm M} = \pm \mu_{\rm n} B$. One spin polarization is attracted by a strong magnetic field (high-field seekers, "+"), while the other is repelled (low-field seekers, "-").

The above interactions allow efficient transport of UCN through guides to an experiment, long observation times, and close to full polarization of the neutron spins, important ingredients for a highly sensitive EDM measurement.
Gravity, with its potential of $102.5\,$neV per meter elevation also influences ultracold neutrons significantly.

\section{Principles}
\label{principles}
\begin{figure}[h]
\centering
    \includegraphics[page=1,width=\textwidth,  trim = {0.8cm 0.3cm 0.2cm 1.8cm},clip]{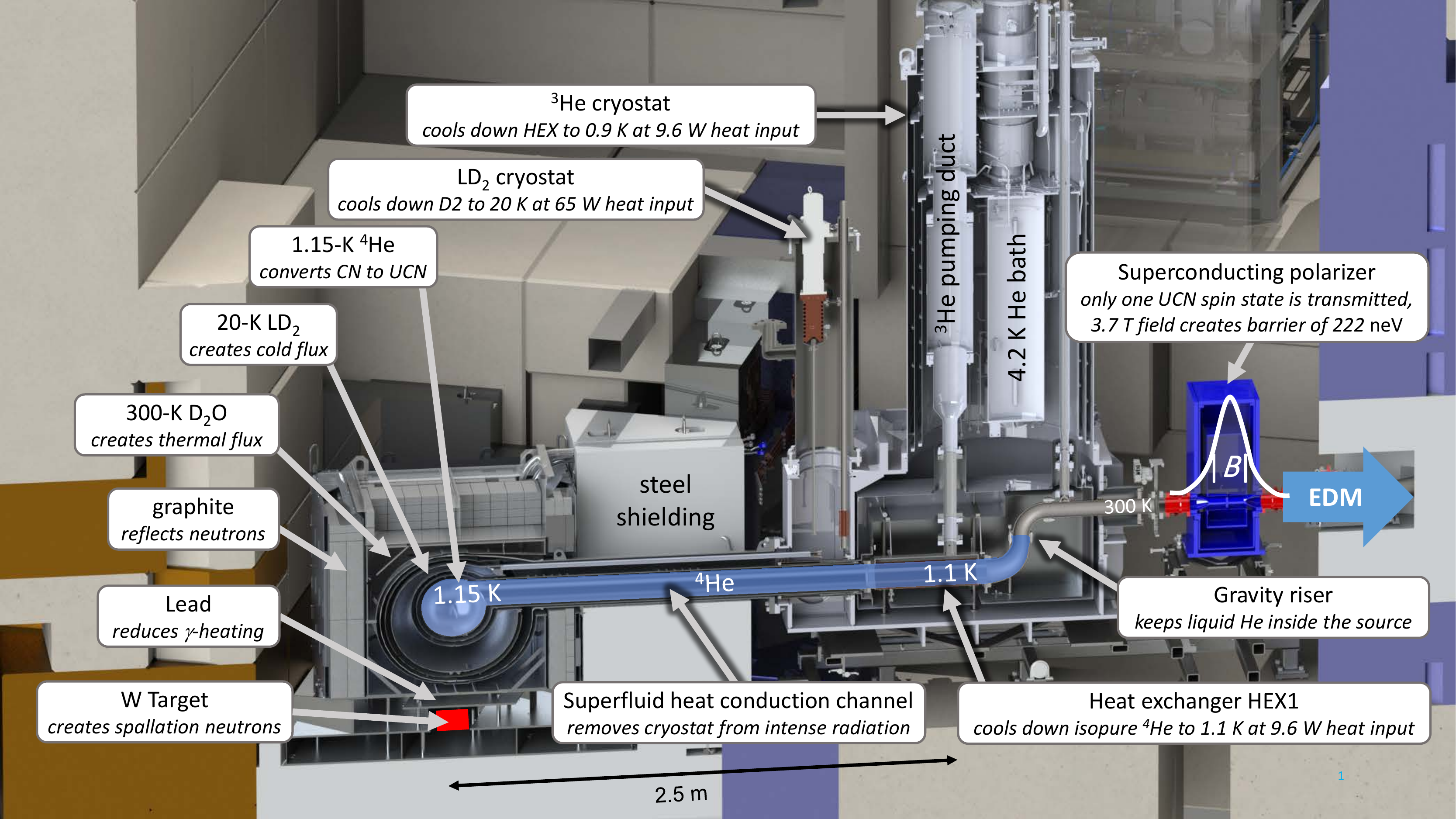}
\caption{Cutaway view of the 3D model for the TUCAN source.}
\label{fig:source}       
\end{figure}

We will now describe the working principles of the UCN source (Fig.~\ref{fig:source}) and EDM apparatus (Fig.~\ref{fig:EDMspectra}) under construction.
\subsection{UCN production}
\label{UCN}
483~MeV protons from the TRIUMF main cyclotron hit a tungsten spallation target to produce fast neutrons~\cite{BL1UNIMA}.
They are moderated in room-temperature heavy water, graphite, and liquid deuterium at 20~K, as shown in Fig.~\ref{fig:source}.
These moderators surround a vessel filled with superfluid, isotopically-purified helium-4 at around 1.1~K.
The geometry has been optimized to create a large flux of cold neutrons inside the helium~\cite{schreyer2020optimizing}.
Here, cold neutrons can excite phonons and rotons, losing almost all their energy in the process to become ultracold.
We estimate a total UCN production rate of $1.4 \times 10^{7} \,$UCN/s at a proton current of $40\,\mu$A~\cite{schreyer2020optimizing}.
The reverse scattering, where a UCN gains momentum by scattering on existing rotons and phonons, is suppressed by cooling the medium.
Thank to this suppression and the fact that pure helium-4 does not absorb neutrons, ultracold neutrons can be stored in the source for tens of seconds, which is similar to the time needed to fill neutrons into experiments such as the nEDM.

\subsection{Cryogenics and UCN extraction}
A total heat load of up to 9.6~W, 8.1~W from the spallation target and 1.5~W estimated static heat load, is deposited in the superfluid helium. 
This heat is transported in a helium-filled channel of 2.5~m length through steel shielding to a large helium-4 to helium-3 heat exchanger (HEX 1).

The smooth surface on the helium-4 side of HEX 1 reduces UCN losses and has an area of $0.26\,$m$^2$.
Vertical fins increase the surface on the helium-3 side to $0.86\,$m$^2$.
An array of pumps connected via large diameter pumping ducts lower the saturated vapor pressure above the liquid helium-3 to 6.6~mbar corresponding to a temperature of 0.89~K. 
The helium-3 is pre-cooled and liquefied in two liquid-helium baths and several heat exchangers~\cite{Okamura_2020}.
The superfluid helium-4  is contained inside the source by an upwards kink in the UCN guide (gravity riser in Fig.~\ref{fig:source}).\footnote{We have estimated the heat load from superfluid film flow to be less than 70~mW and therefore are not planning a knife edge or film burner in this location.}

Accounting for the helium-3 boiling curve from~\cite{MAEDA2000713, TANAKA1989203} and the Kapitza conductivity at the helium-4 interface we measured at KEK~\cite{kawasaki2019cryogenic},
we estimate the helium-4 temperature inside HEX1 to be 1.1~K. 
Liquid helium at this temperature is a two-phase fluid (He-II) with a superfluid and a normal fluid component.
For heat loads as large as in our case, the heat transport is likely limited by turbulent counterflow between the two fluid components, as described by Gorter and Mellink~\cite{GORTER1949285}.
\begin{figure}[h]
\centering
    \includegraphics[page=1,width=0.7\textwidth,  trim = {1.5cm 2.2cm 1.7cm 2.2cm},clip]{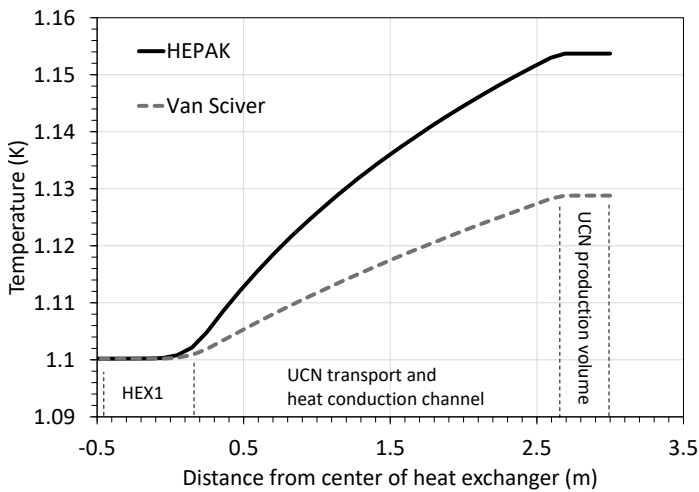}
\caption{Temperature profile inside the He-II of the heat conduction channel in the TRIUMF UCN source according to two parametrizations of Gorter-Mellink heat transport: Van Sciver~\cite{van2012helium} and HEPAK~\cite{hepak2005}}
\label{Tgradient}       
\end{figure}
We have calculated the resulting temperature gradient for two parametrizations of the Gorter-Mellink heat conductivity.
Results using Van Sciver's formalism~\cite{van2012helium} and data from HEPAK~ \cite{hepak2005} are shown in Fig.~\ref{Tgradient}.
From the less favorable HEPAK data, we estimate an average upscattering lifetime for the UCN of 29~s in the source from equation $\tau_{\rm up}^{-1} (T) = B T^7$.
Our conservative estimate of $B = 0.016\,$s$^{-1}$K$^{-7}$ stems from prior measurements~\cite{leung2016ultracold,golub1983operation, YoshikiPhysRevLett1992} and our own extraction based on experiments with a prototype UCN source using superfluid helium~\cite{svanb2023}.
We only account for two-phonon scattering and assume the single-phonon and roton-phonon scattering contributions are negligible.
Upscattering in the helium vapour above the liquid has a minor impact, reducing the lifetime of the neutrons by around 0.1~s.

UCN are extracted through an aluminum foil ($U_{\rm Al} = 54\,$neV) at room temperature separating the helium-vapour-filled region above the gravity riser from the vacuum guides to the experiment. It is mounted inside the warm bore of a superconducting magnet that provides a 3.7~T field ($U_{\rm M} = 222\,$neV).
This accelerates high-field-seeking UCN through the foil ($U_{\rm Al} - U_{\rm M}  = -168\,$neV) and reflects low-field-seeking UCN below $U_{\rm Al} + U_{\rm M} = 276\,$neV, creating a fully polarized UCN flow to the EDM experiment.

Cooling of the liquid-deuterium moderator to around 20~K at a heat load of 65~W from the spallation target is provided by a Gifford-McMahon cryo-cooler inside the LD$_2$ cryostat.
It is connected to the deuterium moderator volume via a thermosyphon loop.

\subsection{EDM experiment}
\label{sec:EDM}
\begin{figure}[h]
\centering
    \includegraphics[page=2,width=0.59\textwidth,  trim = {8.5cm 0cm 3.2cm 0cm},clip]{Figures.pdf}
     \includegraphics[width=0.40\textwidth,  trim = {0cm 0.5cm 10cm 0.9cm},clip]{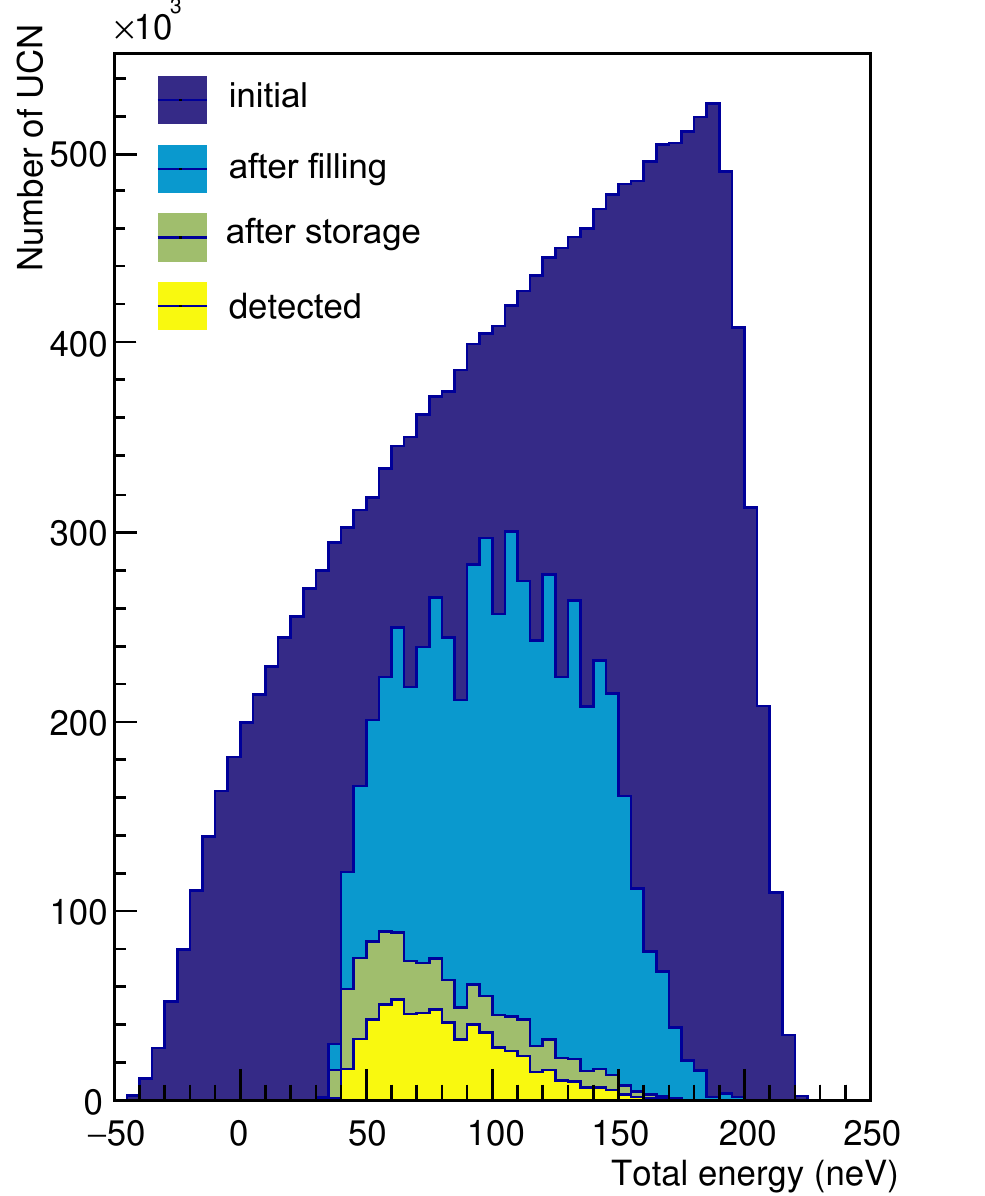}
\caption{\textit{Left}: Cutaway view of the 3D model for the TUCAN EDM Apparatus. (A) 4-layer magnetically shielded room, (B) Self-shielded $B_0$ coil, (C) EDM cell valve actuation, (D) EDM cell shutter, (E) Central high-voltage electrode, (F) UCN guide, (G) High-voltage feeder, (H) High-voltage vacuum feed through, (I) EDM detector switch, (J) UCN and vacuum valve, (K) Spin analyzers and detectors, (L) Coil compensating the ambient magnetic field, (M) Support structure. \textit{Right}: Energy spectra of ultracold neutrons determined by Monte Carlo simulations using PENTrack~\cite{schreyer2017pentrack}. The initial spectrum produced in the source is scaled by a factor of 0.01 in the plot.}
\label{fig:EDMspectra}       
\end{figure}
The TUCAN EDM experiment will use Ramsey's technique of separated oscillating fields~\cite{RamseyMethod} in a double-chamber EDM spectrometer at room temperature as shown in Fig.~\ref{fig:EDMspectra}.
This is comparable to other current EDM experimental efforts~
\cite{PhysRevC.97.012501, ayres2021design, wurm2021panedm}.
Polarized ultracold neutrons will be filled into two vertically stacked EDM cells, each with a height of 0.16~m and a diameter of 0.5~m.
A central high-voltage electrode at 200~kV in between the two cells and two ground electrodes create vertical electric fields with average magnitudes of $E = \pm 12.5\,$kV/cm in top and bottom cell, respectively.\footnote{Our simulations conservatively assume NiP coated electrodes, but we intend to use diamond-like carbon (DLC) coating due to its better performance.}
Insulating rings coated with deuterated polystyrene (dPS) form the outer wall between the electrodes.
Non-magnetic, linearly actuated shutters complete the EDM cells.

The required homogeneous and stable magnetic field environment is created by a magnetically shielded room with 4 layers of mu-metal for DC shielding and one copper layer for AC shielding.
The estimated DC shielding factor is $\approx 10^5$.
A self-shielded coil and several shim coils create a uniform, vertical magnetic field on the order of $B_0 = 1\,\mu$T.
We will introduce polarized mercury vapour into the EDM cells to probe the magnetic field in the same volume that the UCN occupy. 
The $\pi$/2 spin flip required for both UCN and Hg will be initiated by a magnetic field of roughly 10~nT that oscillates close to the Larmor precession frequency of either species.

After the Ramsey cycle, four Li-glass scintillation detectors~\cite{jamiesonLi2017}, two connected to each EDM cell, count the neutrons. 
A spin flipper and a spin analyzer using a magnetized iron film coated onto an aluminum foil are placed on top of each detector.
Low field seekers are not able to penetrate the iron film ($U_{\rm Fe} = 209\,$neV) since the potential barrier is increased to $U_{\rm Fe} - \mu_{\rm n} B_{\rm Fe} = 329.6\,\mathrm{neV}$ by the magnetic field $B_{\rm Fe} = 2$\,T inside the saturated iron film.
High-field seekers only see a barrier of $U_{\rm Fe} + \mu_{\rm n} B_{\rm Fe} = 88.4\,\mathrm{neV}$ that is overcome by placing the analyzers low enough so that the UCN gain more energy in the gravitational fields.
The spin-flipping AC magnetic fields are only activated on one arm per EDM cell, therefore allowing simultaneous counting of both polarization states of the UCN for both cells.

\section{Performance estimation}
\label{calc}
To estimate and improve the UCN source and EDM experiment performance we studied its UCN particle and spin transport characteristics in great detail.
Many factors have an influence. \\
\textbf{Absorption and upscattering} on and in materials can be expressed via an imaginary Fermi potential $U_{\rm F} = V_{\rm F} - i W_{\rm F}$.
The larger the real part $V_{\rm F}$ of wall materials, the larger the energy spread of UCN may be;
the larger the imaginary part $W_{\rm F}$ the higher the loss probability per wall bounce, see~\cite{schreyer2017pentrack}. \\
The \textbf{geometry} of UCN storage vessels and guides influences the mean free path between wall hits and therefore the losses of UCN due to $W_{\rm F}$.
Any elevation change of UCN transport components leads to a kinetic energy change;
some neutrons might not reach higher elevations. \\
 The majority of UCN wall bounces is specular, but roughness of materials causes \textbf{non-specular reflection}.
The simplest model called Lambert reflection assumes a cosine distribution of the outgoing neutron directions around the surface normal for any angle of incidence~\cite{lambert1760photometria}.
In this model, the Lambert reflection probability $P_{\rm L}$ is the sole input parameter. \\
Wall interactions of UCN can also cause \textbf{spin flips}, important for any experiment requiring polarized neutrons. This can be expressed by a depolarization probability per wall bounce $\beta$. \\
\textbf{Magnetic fields} influence UCN trajectories via forces in magnetic gradient fields.
The precession of neutron spins in magnetic fields is a very important aspect for the neutron EDM experiment.

All of these interactions of ultracold neutrons are included in the Monte Carlo simulation package PENTrack~\cite{schreyer2017pentrack}.
PENTrack performs spatial tracking of UCN in gravitational and magnetic fields taking into account surface interactions described above as well as spin tracking in magnetic and electric fields.
We used this package for extensive studies of many aspects of the TUCAN components and their influence on the yield of the UCN source and the sensitivity of the EDM experiment~\cite{stevethesis}. 

\subsection{Input parameters and simulation procedure}
\begin{figure}[h]
\centering
\sidecaption
    \includegraphics[width=\textwidth,  trim = {0cm 7cm 3cm 8cm},clip]{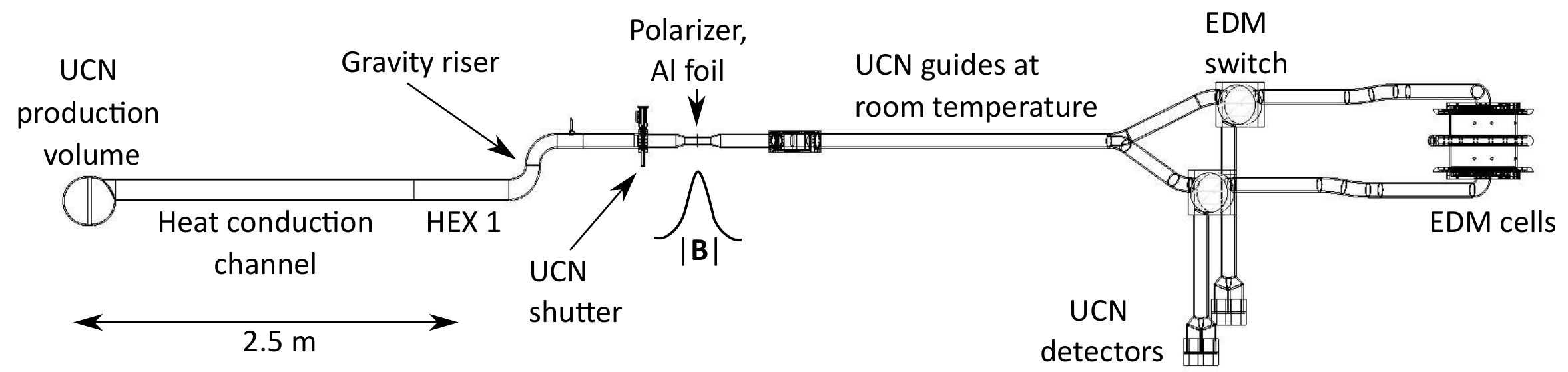}
\caption{Side view of one exemplary geometry used to simulate the TUCAN source and EDM experiment.}
\label{fig:simmodel}       
\end{figure}
Tab.~\ref{tab:input} provides a summary of input parameters used in our simulations and calculations.
We have run sensitivity studies on most parameters and on a lot of geometrical dimensions of the UCN source and EDM experiment.
These studies can be found in~\cite{stevethesis}.

In this work, due to space constraints, we concentrate on the UCN source as it is being built and the EDM experiment as currently planned, see Figs.~\ref{fig:source} and \ref{fig:EDMspectra}. The goal is to estimate the required measurement time to achieve a statistical sensitivity of $10^{-27}\,e$cm (1$\sigma$).

\begin{table}
\centering
\caption{Selected input parameters for UCN Monte Carlo simulations and the sensitivity calculation.}
\label{tab:input}       
\begin{tabularx}{\textwidth}{XlX}
\hline
Parameter & Value & Source  \\\hline
Kinetic energy spectrum produced in source & $\propto \sqrt{E}$, Fig.~\ref{fig:EDMspectra} & Low-energy tail of Maxwell spectrum \\
Base temperature of superfluid helium at HEX1 & 1.1~K & Cryogenic calculations of cryostat~\cite{Okamura_2020}, 3He boiling curve~\cite{MAEDA2000713, TANAKA1989203}, Kapitza resistance~\cite{kawasaki2019cryogenic}  \\
Temperature gradient in superfluid helium & Fig.~\ref{Tgradient} & Gorter-Mellink formalism~\cite{van2012helium, hepak2005} \\
Upscattering lifetime in liquid helium $\tau_{\rm up}^{-1} (T) = B T^7$ & $B = 0.016\,$1/s/K$^7$ & \cite{leung2016ultracold,golub1983operation, YoshikiPhysRevLett1992,svanb2023} \\
\multicolumn{3}{l}{Fermi potential of} \\
- superfluid helium & $19\,\mathrm{neV} - i \frac{\hbar}{2} B T^7$ & \cite{leung2016ultracold,golub1983operation, YoshikiPhysRevLett1992,svanb2023} \\
- NiP coated components & $213 - i 0.07\,$neV & UCN experiments at TRIUMF, \\
- dPS insulator ring & $171 - i 0. 047\,$neV & J-Parc, and LANL~\cite{ahmed2018first, NiPchar2022} \\
\multicolumn{3}{l}{Lambert diffuse reflection  probability in} \\
- cryogenic region & $P_L = 0.15$ & Roughness measurements \\
- room temperature guides & $P_L = 0.03$ & UCN expt's at TRIUMF~\cite{ahmed2018first} \\
Spin-flip prob. per wall bounce & $3 \times 10^{-5}$ & $10 \times$ higher than \cite{Tang:2015pca, brys2005measurement, bondar2017losses} \\ 
Avg. electric field in EDM cells & 12.5~kV/cm & Finite element simulations \\
Total spin coherence lifetime $T_2$ & $> 800~s$ & Estimate based on \cite{PhysRevLett.97.131801,Abel:2020gbr} \\
\hline
\end{tabularx}
\end{table}

In the simulation, we separate the EDM experimental cycle into four major stages. 
(1) UCN production in the source: For a duration $T_{\rm P}$ UCN are produced with a homogeneous spatial distribution and uniform starting angle distribution inside the UCN production volume, see Fig.~\ref{fig:simmodel}. The UCN shutter is closed. 
(2) UCN production and filling of the EDM cells: the UCN shutter and EDM cell shutters are opened, UCN production continues and the neutrons are allowed to diffuse into the EDM cells for a duration $T_{\rm F}$. The spins are filtered by the magnetic field of the polarizer. 
(3) UCN storage and Ramsey period: UCN production stops, the EDM cell shutter is closed, UCN are stored inside the cells. A first spin flip starts free precession of the UCN spins. After the free precession time $T_{\rm R}$, a second spin flip turns any change of the Larmor precession frequency into a polarization change (Ramsey method). 
(4) UCN detection and spin analysis: The EDM cell shutter is opened, UCN can diffuse to spin-sensitive detectors for a duration $T_{\rm det}$, allowing for a polarization measurement. 

\noindent
This separation allows us to find the durations for all stages that minimize the measurement time required to reach the desired sensitivity, see~\cite{stevethesis}.

The magnetic field of the superconducting polarizer is included in the simulation as a 3D field map, created by an Opera finite-element simulation~\cite{opera} and cross checked by magnetic field measurements of the actual magnet.
The spin analyzers and detectors are realized slightly differently in our simulation model.
Spin flippers are represented by volumes inside the UCN guides with a spin flip probability of 99\%, our anticipated spin-flipper efficiency.
The highly localized magnetic field inside the very thin, iron-coated aluminum foil of a spin analyzer cannot be represented via a field map.
The magnetization is instead represented by a spin-dependent pseudo potential inside the foil as described at the end of Sec.~\ref{sec:EDM}.

To determine the sensitivity reach of the EDM experiment we use the well-known sensitivity formula for Ramsey-type EDM experiments
$
        \sigma(d_n)  \geq \frac{\hbar}{2E\alpha_{\rm det} T_{\rm R}\sqrt{N_\mathrm{det}}}.
$
$E$ denotes the electric field, $\alpha_{\rm det}$ the visibility of the Ramsey pattern, mainly driven by polarization, depolarization and spin coherence. 
$T_{\rm R}$ is the free precision time and $N_\mathrm{det}$ the number of UCN detected per Ramsey cycle.

We expect to have a magnetically quiet-enough environment at TRIUMF to conduct the EDM experiment outside business hours (18:00 to 7:00 on weekdays and the whole weekend).
Our figure of merit is the number of days required to reach a statistical sensitivity of $10^{-27}\,e$cm (1$\sigma$) when taking data during these hours.

\subsection{Results}
\begin{figure}[h]
\centering
    \includegraphics[page=1,width=0.49\textwidth,  trim = {0.2cm 0.3cm 0.2cm 0.3cm},clip]{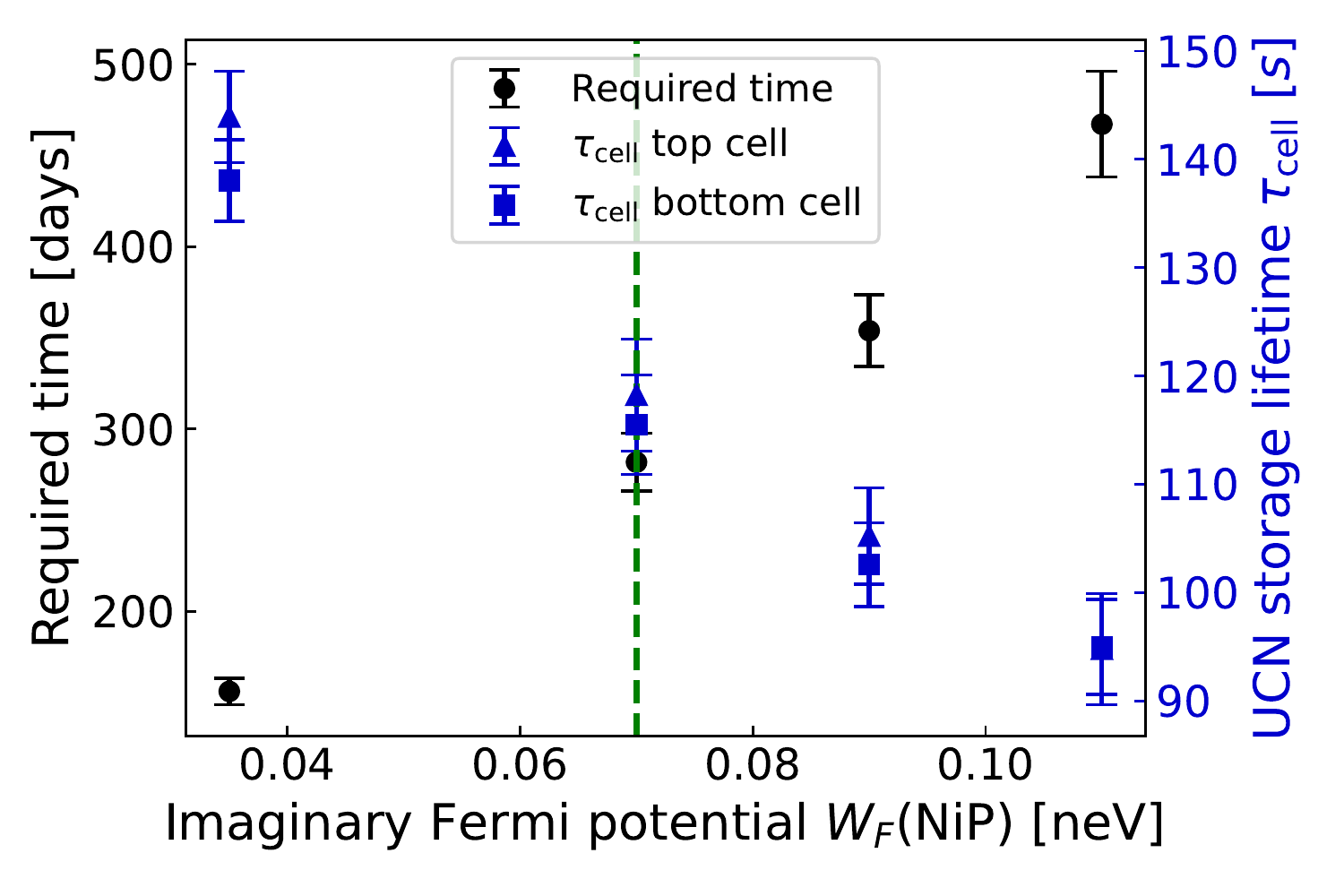}
    \includegraphics[page=1,width=0.49\textwidth,  trim = {0.2cm 0.3cm 0.2cm 0.3cm},clip]{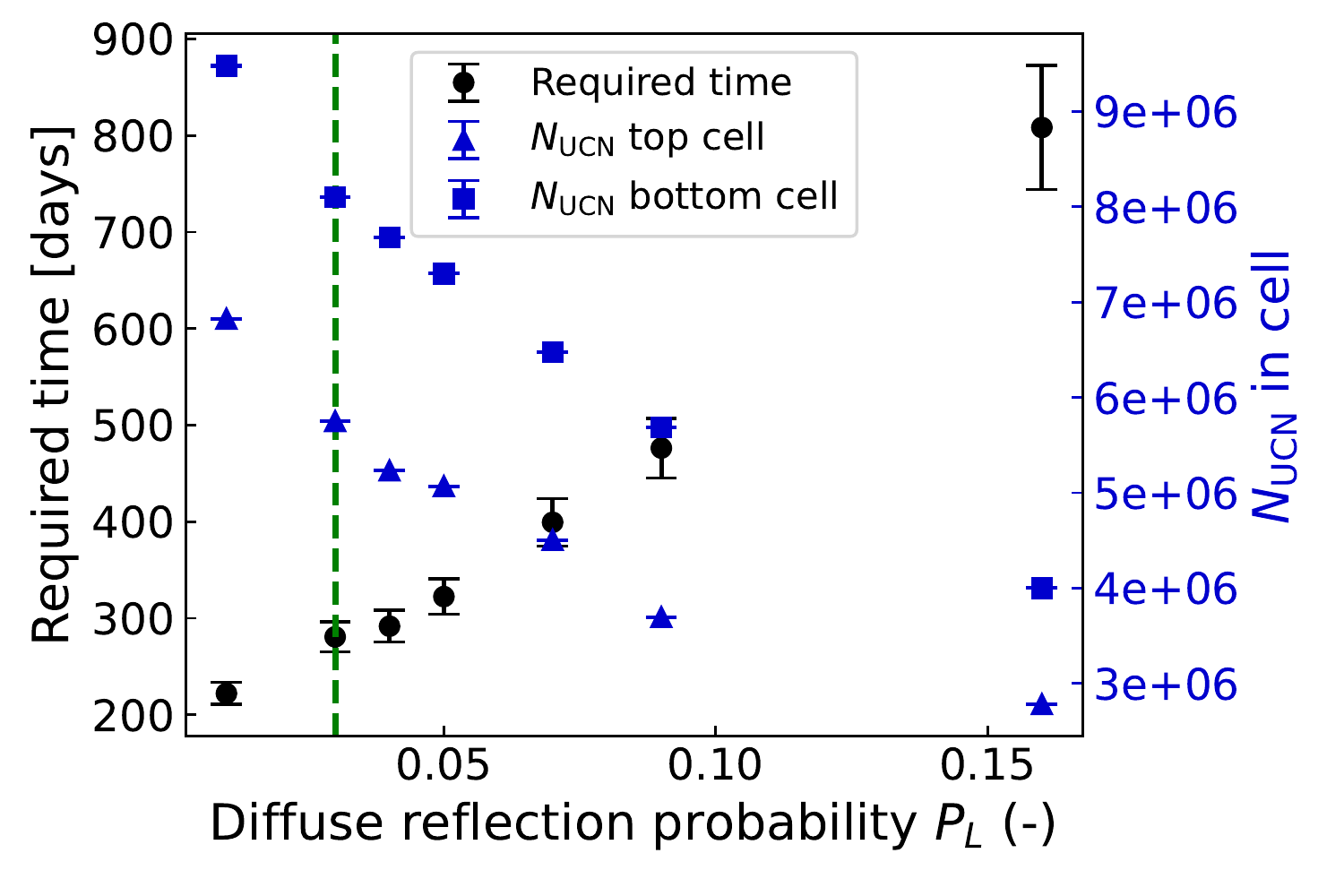}
    \includegraphics[page=1,width=0.49\textwidth,  trim = {0.2cm 0.4cm 0.2cm 0.3cm},clip]{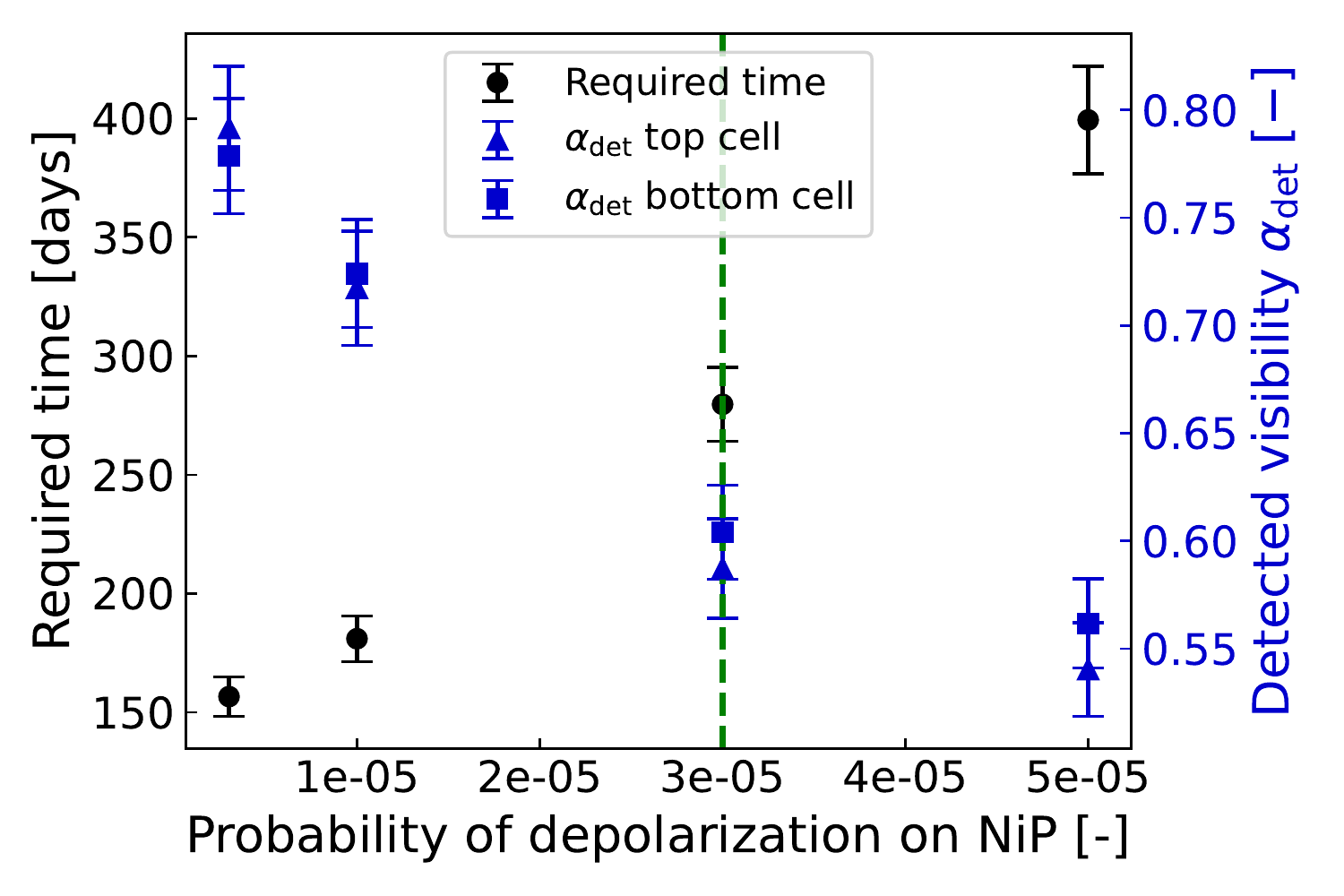}
    \includegraphics[page=1,width=0.49\textwidth,  trim = {0.2cm 0.4cm 0.2cm 0.3cm},clip]{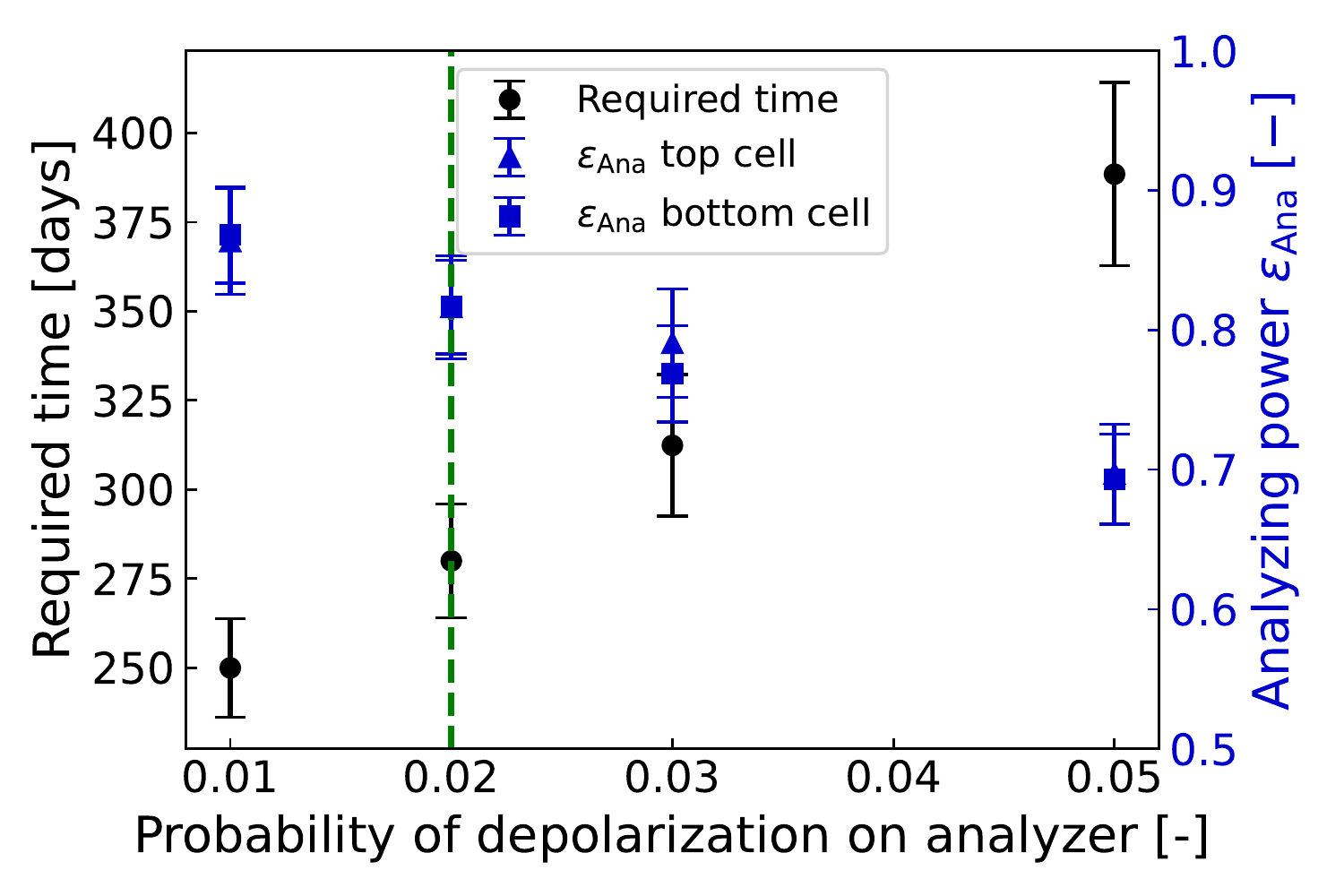}  
    
\caption{Four PENTrack sensitivity studies showing the required measurement time to reach a statistical sensitivity of $10^{-27}~e$cm ($1\sigma$) and a selected other parameter for each plot. Our parameter assumptions are indicated by the vertical green lines. For more information, please refer to the text.
}
\label{fig:scans}       
\end{figure}

Extensive simulation campaigns have been performed to understand and improve the performance of the TUCAN source and EDM experiment.
The dependence of the results on the assumptions made, especially material properties, were studied. 
We will mainly focus on dependence on three parameters very relevant to UCN storage and transport that are notoriously hard to measure: the imaginary Fermi potential, the diffuse reflection probability in the Lambert model and the spin flip probability on wall interaction.

In the simulation we assume all UCN exposed surfaces to be coated with electro-less nickel (NiP) with a high phosphorous content above 10\%, except the Al foil inside the polarizer ($54.1 - i 0.00281\,$neV), the Al analyzer foil coated with iron ($209.1 - i0.000872\,$neV) and the EDM cell insulators coated with dPS ($171 - i0.047\,$neV).
Our measurements indicate a range for the NiP imaginary Fermi potential of roughly [0.03,0.09]~neV~\cite{NiPchar2022}.
Fig.~\ref{fig:scans} (top left) indicates that the required measurement time to reach $10^{-27}\,e$cm varies by a factor of three mainly driven by the UCN storage lifetime in the EDM cells.

The diffuse reflection probability mainly impedes UCN transport through guides. 
Varying it from 1~\% to 16~\% changes the number of neutrons loaded into the EDM cells and the measurement time dramatically, see Fig.~\ref{fig:scans} (right top).

For spin-sensitive experiments like the EDM, spin transport properties are very important. 
We set the baseline probability for spin flip on wall bounce on the NiP surfaces to a very conservative $3 \times 10^{-5}$, 10 times higher than literature values~\cite{Tang:2015pca, brys2005measurement, bondar2017losses}.
A parameter scan in our simulation shows lower depolarization at the walls would reduce the measurement time and improve the visibility $\alpha_{\rm det}$ of the Ramsey fringes significantly.

Additional depolarization is expected at the iron coated analyzer foil. 
Varying it from 1\% to 5\% has a rather mild effect on the required measurement time and changes the analyzing power of the spin sensitive detection by around 25\%.

We performed numerous simulation studies varying geometrical parameters of source and experiment, leading to the design as shown in Fig.~\ref{fig:simmodel}~\cite{stevethesis}.
The optimum experiment timings are: $T_{\rm P} = 20\,$s, $T_{\rm F} = 105\,$s, $T_{\rm R} = 188\,$s, and $T_{\rm det} = 49\,$s.
The UCN storage lifetime inside the source, an important benchmark parameter, is estimated to be $19.0\pm0.2$~s, mainly dominated by the upscattering lifetime inside the liquid helium.
The expected storage lifetimes inside the top and bottom EDM cells are $119\pm4$~s and $116\pm4$~s, respectively.
The initial energy spectrum as well as spectra right after stage (2) filling, (3) storage and for the detected UCN are depicted in Fig.~\ref{fig:EDMspectra}.
Higher-energy UCN are less likely to survive and to be detected, favoring an optimization towards lower energy UCN.

For our conservative baseline assumptions, we expect to fill $(1.38\pm0.02) \times 10^{7}$ neutrons into the EDM experiment, which corresponds to a density of $217\pm3\,$UCN/cm$^3$, and detect $N_\mathrm{D} = (1.43\pm0.02) \times 10^6$ after the Ramsey cycle assuming a detector efficiency of 90\%.
The initial UCN polarization in the EDM cell is projected to be $(92\pm2)$\%  taking into account the depolarization per bounce from Tab.~\ref{tab:input}.
This reduces to $(73\pm2)$\% after the Ramsey cycle, corresponding to a spin coherence time $T_2 \approx 830\,$s.
The detected visibility is estimated to be $\alpha_{\rm det} = (60 \pm 2)$\%.
The sensitivity per cycle then becomes $(1.94\pm0.06) \times 10^{-25}\,e$cm.

This should allow us to achieve a statistical sensitivity of $10^{-27}\,e$cm (1$\sigma$) in $281\pm16$ calendar days of data taking.
Quoted uncertainties are calculated from the simulation statistics and fitting errors.
We will perform commissioning experiments to validate the input parameters for our calculations as we build the source and experiment components.

\section{Conclusion}
The TUCAN collaboration is building an ultracold neutron source based on a spallation target and a superfluid helium converter as well as a room-temperature apparatus to search for the electric dipole moment of the neutron at TRIUMF.
Extensive simulation campaigns have been performed to understand and improve source and experiment performance.
Their results have driven the designs of UCN source and EDM apparatus as shown above.

We estimate to reach a statistical sensitivity of  $10^{-27}\,e$cm (1$\sigma$) in $281\pm16$ calendar days or less than three years of data taking.

\bibliography{references}

\end{document}